\documentclass[journal=jcisd8, manuscript=article]{achemso}
\usepackage{amsmath, amsfonts, amssymb, graphicx, subfigure, xcolor, soul,indentfirst,siunitx,booktabs,multicol,blindtext}
\usepackage{multirow,makecell}
\usepackage[labelfont=bf]{caption}

\author{Hao Tian}
\affiliation{Department of Chemistry, Center for Research Computing, Center for Drug Discovery, Design, and Delivery (CD4), Southern Methodist University, Dallas, Texas, United States of America}

\author{Rajas Ketkar}
\affiliation{Wakeland High School, Frisco, Texas, United States of America}

\author{Peng Tao}
\affiliation{Department of Chemistry, Center for Research Computing, Center for Drug Discovery, Design, and Delivery (CD4), Southern Methodist University, Dallas, Texas, United States of America}
\email{ptao@smu.edu}

\title{Accurate ADMET Prediction with XGBoost}

\begin{document}

\clearpage

\begin{abstract}

The absorption, distribution, metabolism, excretion, and toxicity (ADMET) properties are important in drug discovery as they define efficacy and safety. In this work, we applied an ensemble of features, including fingerprints and descriptors, and a tree-based machine learning model, extreme gradient boosting, for accurate ADMET prediction. Our model performs well in the Therapeutics Data Commons ADMET benchmark group. For 22 tasks, our model is ranked first in 18 tasks and top 3 in 21 tasks. The trained machine learning models are integrated in ADMETboost, a web server that is publicly available at \url{https://ai-druglab.smu.edu/admet}.

\end{abstract}

\section{Introduction}

Properties such as absorption, distribution, metabolism, excretion, and toxicity (ADMET) are important in small molecule drug discovery and therapeutics. It is reported that many clinical trials fail due to the deficiencies in ADMET properties. \cite{kola2004can,kennedy1997managing} While profiling ADMET in the early stage of drug discovery is desirable, experimental evaluation of ADMET properties is costly with limited available data. 

Recent developments in machine learning (ML) promote research in chemistry and biology \cite{song2020unraveling,zhang2020multiscale,tian2021explore} and bring new opportunities for ADMET prediction. ADMETLab \cite{dong2018admetlab} provides 31 ADMET endpoints with six machine learning models, and further advanced to 53 endpoints using a multi-task graph attention network \cite{xiong2021admetlab}. vNN \cite{schyman2017vnn} is a web server that applies variable nearest neighborhood to predict 15 ADMET properties. admetSAR \cite{cheng2012admetsar} and admetSAR 2.0 \cite{yang2019admetsar} are also ML based web servers for drug discovery or environmental risk assessment with random forest, support vector machine and k-nearest neighbors models. As a fingerprint-based random forest model, FP-ADMET \cite{venkatraman2021fp} evaluated over 50 ADMET and ADMET-related tasks. In these ML models, small molecules are provided in SMILES representations and further featurized using fingerprints, such as extended connectivity fingerprints \cite{rogers2010extended} and Molecular ACCess System (MACCS) fingerprints \cite{durant2002reoptimization}. Beside these, there are many other fingerprints and descriptors that can be used to for ADMET prediction, such as PubChem fingerprints and Mordred descriptors. Taking advantage of all possible features enables sufficient learning process for machine learning models. 

One common issue is that many machine learning models in previous work are trained on different datasets, which leads to unfair comparison and evaluation of ML models. As a curated dataset, Therapeutics Data Commons (TDC) \cite{Huang2021tdc} unifies resources in therapeutics for systematic access and evaluation. There are 22 tasks in TDC ADMET benchmark group, each with small molecules SMILES representations and corresponding ADMET property values or labels. 

Extreme gradient boosting (XGBoost) \cite{chen2016xgboost}, developed by Chen \textit{et al.}, is a powerful machine learning model and has been shown to be effective in regression and classification tasks in biology and chemistry \cite{chen2020improving,tian2020deciphering,tian2021passer,deng2021xgraphboost}. In this work, we applied XGBoost to learn a feature ensemble, including multiple fingerprints and descriptors, for accurate ADMET prediction. Our model performs well in the TDC ADMET benchmark group with 11 tasks ranked first and 19 tasks ranked top 3.

\section{Methods}

\subsection*{Therapeutics Data Commons}

Therapeutics Data Commons is a Python library with an open-science initiative. It holds many therapeutics tasks and datasets including target discovery, activity modeling, efficacy and safety, and manufacturing. TDC provides a unified and meaningful benchmark for fair comparison between different machine learning models. For each ADMET prediction task, TDC splits the dataset into the predefined 80\% training set and 20\% test set with scaffold split, which simulates the real-world application scenario where a well-trained machine learning model would be asked to predict ADMET properties on unseen and structurally different drugs. 

\subsection*{Fingerprints and Descriptors}

Six featurizers from DeepChem \cite{ramsundar2019deep} were used to compute fingerprints and descriptors:

\begin{itemize}
    \item MACCS fingerprints are common structural keys that compute a binary string based on a molecule's structural features.
    \item Extended connectivity fingerprints compute a bit vector by breaking up a molecule into circular neighborhoods. They are widely used for structure-activity modeling.
    \item Mol2Vec fingerprints \cite{jaeger2018mol2vec} create vector representations of molecules based on an unsupervised machine learning approach.
    \item PubChem fingerprints consist of 881 structural keys that cover a wide range of substructures and features. It is used by PubChem for similarity searching. 
    \item Mordred descriptors \cite{moriwaki2018mordred} calculate a set of chemical descriptors such as the count of aromatic atoms or of all halogen atoms.
    \item RDKit descriptors calculate a set of chemical descriptors such as molecular weight and the number of radical electrons.
\end{itemize}

\subsection*{Extreme Gradient Boosting}

Extreme gradient boosting is a powerful machine learning model. It boosts model performance through ensemble that includes decision tree models trained in sequence. 

Let $D = \{ (x_i, y_i) (|D| = n, x_i \in R^m, y_i \in R^n) \}$ represents a training set with $m$ features and $n$ labels. The $j$-th decision tree in XGBoost model makes a prediction for sample $(x_i, y_i)$ by $g_j(x_i) = w_q(x_i)$ where $w_q$ is the leaf weights. The final prediction of XGBoost model is the sum of all $M$ decision tree predictions with $\hat{y_i} = \sum_{j=1}^M g_j(x_i)$. 

The objective function consists of a loss function $l$ and a regularization term $\Omega$ to reduce overfitting: 

\begin{equation}
  \text{obj} (\theta) = \sum_{i=1}^N l(y_i, \hat{y_i}) + \sum_{j = 1}^M \Omega(f_i)
\end{equation}
where $\Omega(f) = \gamma T + \frac{\lambda}{2}\sum_{l=1}^T \omega_l^2$. $T$ represents the number of leaves while $\gamma, \lambda$ are parameters for regularization. 

During training, XGBoost iteratively trains a new decision tree based on the output of the previous tree. The prediction of the $t$-th iteration $\hat{y_i}^{(t)} = \hat{y_i}^{(t-1)} + g_t(x_i)$. The objective function of the $t$-th iteration is:

\begin{equation}
  \text{obj}^{(t)} = \sum_{i=1}^N l(y_i, \hat{y_i}^{(t-1)} + g_t(x_i)) + \Omega(f_i)
\end{equation}

XGBoost introduces first and second derivatives of this objective function, which can be expressed as follows by applying Taylor expansion at second order:

\begin{equation}
\begin{split}
  \text{obj}^{(t)} &\simeq \sum_{i=1}^N [l(y_i, \hat{y_i}^{(t-1)}) + \partial_{\hat{y}^{(t-1)}} l(y_i, \hat{y}^{(t-1)}) f_t(x_i) \\
  &+ \frac{1}{2} \partial_{\hat{y}^{(t-1)}}^2 l(y_i, \hat{y}^{(t-1)}) f_t^2(x_i)] + \Omega(f_i)
\end{split}
\end{equation}

\begin{table}[t]
\centering
\caption{Fine-tuned XGBoost Parameters}
\begin{tabular}{c c c}
\hline
Name and Description & Values\\
\hline
\multirowcell{2}{n\_estimators:\\Number of gradient boosted trees.} & [50, 100, 200, 500, 1000]\\
\\
\multirowcell{2}{max\_depth: \\ Maximum tree depth.} & [3, 4, 5, 6, 7]\\
\\
\multirowcell{2}{learning\_rate: \\ Boosting learning rate.} & [0.01, 0.05, 0.1, 0.2, 0.3]\\
\\
\multirowcell{2}{subsample: \\ Subsample ratio of instances.} &[0.5, 0.6, 0.7, 0.8, 0.9, 1.0]\\
\\
\multirowcell{2}{colsample\_bytree: \\ Subsample ratio of columns.} & [0.5, 0.6, 0.7, 0.8, 0.9, 1.0]\\
\\
\multirowcell{2}{reg\_alpha: \\ L1 regularization weights.} & [0, 0.1, 1, 5, 10]\\
\\
\multirowcell{2}{reg\_lambda: \\ L2 regularization weights.} & [0, 0.1, 1, 5, 10]\\
\\
\hline
\end{tabular}
\label{tab:params}
\end{table}

A total of seven parameters are being fine-tuned with selected value options and are listed in Table \ref{tab:params}. Default values are used for other parameters. 

\subsection*{Performance Criteria}

For regression tasks, mean absolute error (MAE) and Spearman's correlation coefficient are considered to evaluate model performance:

\begin{itemize}
    \item MAE is used to measure the deviation between predictions $y_i$ and real values $x_i$ in $n$ sample size.
    \begin{equation}
        \text{MAE} = \frac{\sum_{i=1}^n |y_i - x_i|} {n}
    \end{equation}

    \item Spearman's correlation coefficient $\rho$ measures the correlation strength between two ranked variables. Where $d_i$ represents the difference in paired ranks, 
    \begin{equation}
        \rho = 1 - \frac{6\sum d_i^2}{n(n^2 - 1)}
    \end{equation}

\end{itemize}

For binary classification tasks, area under curve (AUC) is calculated with receiver operating characteristic (ROC) and precision-recall curve (PRC). For both metrics, a higher value indicates a more powerful model. 
\begin{itemize}
    \item AUROC is the area under the curve where x-axis is false positive rate and y-axis is true positive rate. 
    \item AUPRC is the area under the curve where x-axis is recall and y-axis is precision.
\end{itemize}

All metrics are calculated with evaluation functions provided by the TDC APIs.

\section{Results and Discussion}

\subsection*{Model Performance}

\begin{table*}[!t]
\centering
\footnotesize
\caption{Model Evaluation on the TDC ADMET Leaderboard\textsuperscript{\emph{a}}}
\begin{tabular}{cc|cc|cc}
\toprule
\multicolumn{2}{c|}{TDC} & \multicolumn{2}{c|}{Current Top 1} & \multicolumn{2}{c}{XGBoost} \\
\midrule
Task & Metric & Method & Score & Score & Rank\\
\midrule
\multicolumn{6}{l}{Absorption}\\
\midrule
Caco2 & MAE & RDKit2D + MLP & 0.393 ± 0.024 & 0.288 ± 0.011 & 1st \\
HIA & AUROC & AttrMasking & 0.978 ± 0.006 &  0.987 ± 0.002 & 1st \\
Pgp & AUROC & AttrMasking & 0.929 ± 0.006 & 0.911 ± 0.002 & 4th \\
Bioav & AUROC & RDKit2D + MLP & 0.672 ± 0.021 & 0.700 ± 0.010 & 1st \\ 
Lipo & MAE & ContextPred & 0.535 ± 0.012 & 0.533 ± 0.005 & 1st \\
AqSol & MAE & AttentiveFP & 0.776 ± 0.008 & 0.727 ± 0.004 & 1st \\
\midrule
\multicolumn{6}{l}{Distribution}\\
\midrule
BBB & AUROC & ContextPred & 0.897 ± 0.004 & 0.905 ± 0.001 & 1st \\
PPBR & MAE & NeuralFP & 9.292 ± 0.384 & 8.251 ± 0.115 & 1st \\
VDss & Spearman & RDKit2D + MLP & 0.561 ± 0.025 & 0.612 ± 0.018 & 1st\\
\midrule
\multicolumn{6}{l}{Metabolism}\\
\midrule
CYP2C9 Inhibition & AUPRC & AttentiveFP & 0.749 ± 0.004 & 0.794 ± 0.004 & 1st \\
CYP2D6 Inhibition & AUPRC & AttentiveFP & 0.646 ± 0.014 & 0.721 ± 0.003 & 1st \\
CYP3A4 Inhibition & AUPRC & AttentiveFP & 0.851 ± 0.006 & 0.877 ± 0.002 & 1st \\
CYP2C9 Substrate & AUPRC & Morgan + MLP & 0.380 ± 0.015 & 0.387 ± 0.018 & 1st \\
CYP2D6 Substrate & AUPRC & RDKit2D + MLP & 0.677 ± 0.047 & 0.648 ± 0.023 & 3rd \\
CYP3A4 Substrate & AUPRC & CNN & 0.662 ± 0.031 & 0.680 ± 0.005 & 1st \\
\midrule
\multicolumn{6}{l}{Excretion}\\
\midrule
Half Life & Spearman & Morgan + MLP & 0.329 ± 0.083& 0.396 ± 0.027 & 1st \\
CL-Hepa & Spearman & ContextPred & 0.439 ± 0.026 & 0.420 ± 0.011 & 2nd \\
CL-Micro & Spearman & RDKit2D + MLP & 0.586 ± 0.014 & 0.587 ± 0.006 & 1st \\
\midrule
\multicolumn{6}{l}{Toxicity}\\
\midrule
LD50 & MAE & Morgan + MLP & 0.649 ± 0.019 & 0.602 ± 0.006 & 1st \\
hERG & AUROC & RDKit2D + MLP & 0.841 ± 0.020 & 0.806 ± 0.005 & 3rd \\
Ames & AUROC & AttrMasking & 0.842 ± 0.008 & 0.859 ± 0.002 & 1st \\
DILI & AUROC & AttrMasking & 0.919 ± 0.008 & 0.933 ± 0.011 & 1st \\
\bottomrule
\end{tabular}

\textsuperscript{\emph{a}} Only models that have been evaluated by most of the tasks are considered. 
\label{leaderboard}
\end{table*}

We first used a random seed to split the overall dataset into a training set (80\%) and a test set (20\%). XGBoost model was trained with the training set using 5-fold cross validation (CV). A randomized grid search CV was applied to search hyper parameters. The parameter set with the highest CV score is used, and the model performance is evaluated on the test set. We repeat this process five times with varying random seeds from zero to four following the TDC guideline. The evaluation results are listed on Table \ref{leaderboard}. In each task, there are at least seven other models or featurization methods being compared against, including DeepPurpose {\cite{huang2020deeppurpose}}, AttentiveFP {\cite{xiong2019pushing}}, ContextPred {\cite{hu2019strategies}}, NeuralFP {\cite{lee2021neuralfp}}, AttrMasking {\cite{hu2019strategies}} and graph convolutional network {\cite{kipf2016semi}}. For all 22 tasks, XGBoost is ranked first for 18 and top 3 for 21 out of 22 tasks, demonstrating the success of XGBoost model in predicting ADMET tasks.

The superior prediction results of XGBoost are explainable. As shown in Table \ref{leaderboard}, previously, there are 13 tasks which the top models are trained using descriptors (RDKit 2D + MLP model) or fingerprints (Morgan + MLP model and AttentiveFP). Inspired by this, XGBoost was trained using a combination of fingerprints and descriptors. These featurization methods cover both structural features (MACCS, extended connectivity, Mol2Vec, and PubChem fingerprints) to chemical descriptors (Mordred and RDKit descriptors) for each given SMILES representation. For a specific property prediction task, XGBoost can take the consideration of all possible molecular features, select the best set of them for prediction, while avoiding over-fitting by controlling tree complexity. Together, these would boost the prediction performance of XGBoost to be superior to other models.

\begin{figure*}[!t]
\centering
  \includegraphics[width=0.98\textwidth]{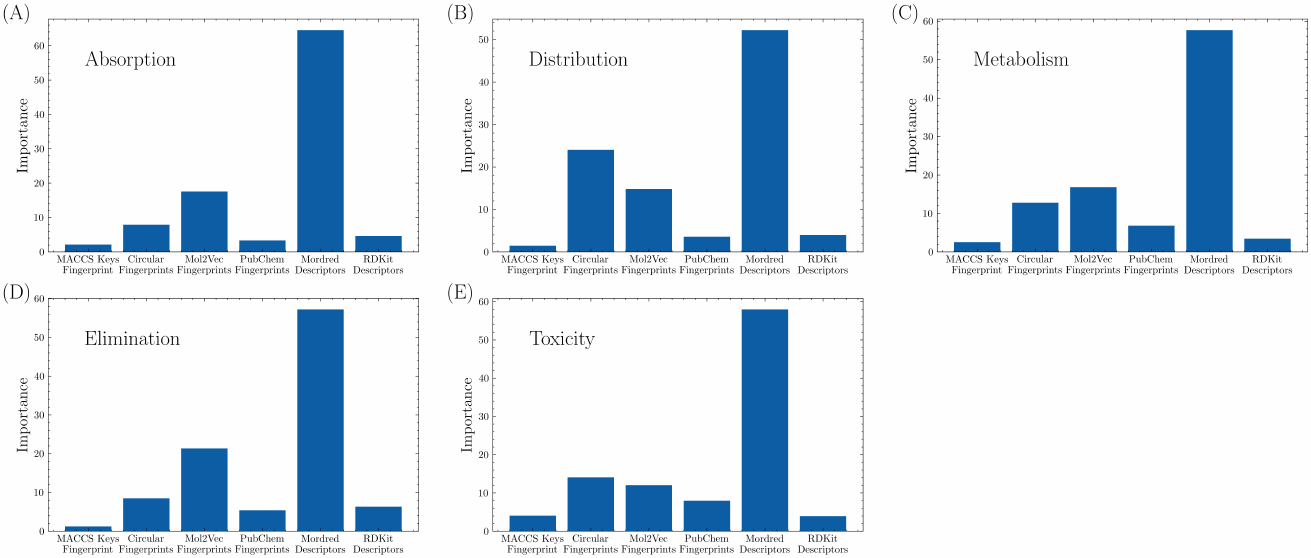}
  \caption{Average feature importance of fingerprints and descriptors in (A) absorption, (B) distribution, (C) metabolism, (D) elimination, and (E) toxicity tasks. }
  \label{fig:importance}
\end{figure*}

\begin{table*}[!t]
\centering
\footnotesize
\caption{Performance comparison of XGBoost models trained with all features and with only Mordred.}
\begin{tabular}{cc|c|c}
\toprule
\multicolumn{2}{c|}{TDC} & XGBoost with all features & XGBoost with Mordred \\
\midrule
Task & Metric & Score & Score\\
\midrule
\multicolumn{4}{l}{Absorption}\\
\midrule
Caco2 & MAE & 0.288 ± 0.011 & 0.301 ± 0.008\\
HIA & AUROC & 0.987 ± 0.002 & 0.990 ± 0.002\\
Pgp & AUROC & 0.911 ± 0.002 & 0.909 ± 0.005\\
Bioav & AUROC & 0.700 ± 0.010 & 0.692 ± 0.016\\ 
Lipo & MAE & 0.533 ± 0.005 & 0.538 ± 0.003\\
AqSol & MAE & 0.727 ± 0.004 & 0.720 ± 0.003\\
\midrule
\multicolumn{4}{l}{Distribution}\\
\midrule
BBB & AUROC & 0.905 ± 0.001 & 0.900 ± 0.001\\
PPBR & MAE & 8.251 ± 0.115 & 7.897 ± 0.061\\
VDss & Spearman & 0.612 ± 0.018 & 0.610 ± 0.005\\
\midrule
\multicolumn{4}{l}{Metabolism}\\
\midrule
CYP2C9 Inhibition & AUPRC & 0.794 ± 0.004 & 0.781 ± 0.002\\
CYP2D6 Inhibition & AUPRC & 0.721 ± 0.003 & 0.694 ± 0.005\\
CYP3A4 Inhibition & AUPRC & 0.877 ± 0.002 & 0.862 ± 0.002\\
CYP2C9 Substrate & AUPRC & 0.387 ± 0.018 & 0.334 ± 0.004\\
CYP2D6 Substrate & AUPRC & 0.648 ± 0.023 & 0.594 ± 0.034\\
CYP3A4 Substrate & AUPRC & 0.680 ± 0.005 & 0.649 ± 0.013\\
\midrule
\multicolumn{4}{l}{Excretion}\\
\midrule
Half Life & Spearman & 0.396 ± 0.027 & 0.373 ± 0.008\\
CL-Hepa & Spearman & 0.420 ± 0.011 & 0.378 ± 0.020\\
CL-Micro & Spearman & 0.587 ± 0.006 & 0.576 ± 0.010\\
\midrule
\multicolumn{4}{l}{Toxicity}\\
\midrule
LD50 & MAE & 0.602 ± 0.006 & 0.602 ± 0.006\\
hERG & AUROC & 0.806 ± 0.005 & 0.763 ± 0.007\\
Ames & AUROC & 0.859 ± 0.002 & 0.856 ± 0.002\\
DILI & AUROC & 0.933 ± 0.011 & 0.928 ± 0.003\\
\bottomrule
\end{tabular}
\label{compare}
\end{table*}

To further understand the importance of each fingerprint and descriptor, for each ADMET task, averaged feature importance is calculated for each feature set and is plotted in Figure \ref{fig:importance}. It is shown that Mordred descriptors are consistently the most important feature in all tasks, followed by Mol2Vec and Circular fingerprints. MACCS Keys fingerprint set is the least important among the five groups of features. As Mordred descriptors are considered significantly more important than other features, we retrained the models in each task using only this feature set. The results are listed in Table \ref{compare}. XGBoost with only Mordred outperformed the base model in three tasks (HIA, Aqsol and PPBR), dropped greatly in Metabolism and Excretion related tasks and was comparable in other tasks.

Searching for the parameter set with best validation performance is necessary. However, there are over 100,000 parameter combinations in the current search space, and it could growth exponentially with additional features being considered. It is challenging to iterate over all possible parameter set to find the best parameter set. In the current study, a randomized grid search CV was used. It should be noted that the randomized grid search CV does not necessarily lead to the global optimum parameter set due to the randomness nature. In recent decades, Bayesian optimization has been developed to search in the hyper parameter space, such as hyperopt \cite{bergstra2013making}, which might be promising under such tasks. 

\subsection*{Web Server}

The trained machine learning models are hosted on the SMU high computing center at \url{https://ai-druglab.smu.edu/admet}. A SMILES representation is required for ADMET predictions. On the result page, molecule structures in both 2D and 3D are displayed using Open Babel \cite{o2011open}. A table is present to summary prediction results under 22 tasks. Each task is colored with green, yellow or red to indicate the optimal level if there is any. The web server has been tested rigorously to respond within seconds. 

\section{Conclusion}

In this study, we applied XGBoost for ADMET prediction. XGBoost can effectively learn molecule features ranging from fingerprints to descriptors. For the 22 tasks on TDC benchmark, our model is ranked first in 11 tasks with all tasks ranked in top 5. The web server, ADMETboost, can be freely accessed at \url{https://ai-druglab.smu.edu/admet}. 

\begin{acknowledgement}
Research reported in this paper was supported by the National Institute of General Medical Sciences of the National Institutes of Health under Award No. R15GM122013. Computational time was generously provided by Southern Methodist University's Center for Research Computing. 
\end{acknowledgement}

\bibliography{ref}

\end{document}